**Mechanism of Tubular Pinch Effect of Dilute Suspension in a Pipe Flow**

Hua-Shu Dou,[1*] Boo Cheong Khoo, [2]

[1]Temasek Laboratories, National University of Singapore, Singapore 117508

[2]Department of Mechanical Engineering, National University of Singapore, Singapore 119260

**\*Email:** tsldh@nus.edu.sg; huashudou@yahoo.com

**Abstract**

Experiments have shown that in dilute suspension flow at laminar state through a circular tube particles migrate towards a concentric annular region with a mean radius of about 0.6 of the tube radius. This phenomenon is well-known as the tubular pinch effect, which is still not fully understood. In this study, the energy gradient method is used to study this phenomenon. It is found that at low Reynolds number particles will move to the position of $r/R=0.58$. Based on the result, the mechanism of this phenomenon is well explained.

PACS numbers:

| | |
|---|---|
| 47.55.Kf | Particle-laden flows |
| 47.61.Jd | Multiphase flows |
| 47.63.Cb | Blood flow in cardiovascular system |
| 47.57.E- | Suspensions |
| 83.80.Hj | Suspensions, dispersions, pastes, slurries, colloids |

## 1. Introduction

The behavior of spheroid particles suspended in a shear flow has been studied extensively for the past half-century since it has a diverse application as in blood flow, micro-separation, chemical processes and others. Segre and Silberberg [1], working with a neutrally buoyant dilute suspension of rigid spheres, were the first to report on the finding of suspensions in laminar flow through a circular tube where it was found that particles tend to migrate towards a concentric annular region with the mean radius of about 0.6 of the tube radius (see also [2]). Subsequently, many follow-on experiments confirmed the same phenomenon (Fig.1) [3-5], and is well-known as the tubular pinch effect [5]. Since creeping flow predicts an absence of lift on the sphere, the phenomenon observed has been largely attributed to the inertia effect [2, 6]. There are several theoretical and simulation works contributed to predict the behaviour of particle migration [6-11], however, the mechanism leading to this phenomenon is still not fully understood with broad consensus.

Recently, Dou [12-13] proposed an energy gradient method with the aim to clarify the phenomenon of flow instability as well as the onset of transition from laminar to turbulent flow applicable to the wall-bounded shear flow. The method has generated agreement with the experimental data in many aspects for plane Poiseuille flow, pipe Poiseuille flow, and plane Couette flow. The energy gradient method has been successfully employed to analyze the Taylor-Couette flow between two concentric rotating cylinders [14].

In this paper, following the proposed principle of energy gradient method, it is shown that the formation of the annulus region of particle concentration in pipe flow is due to the influence of energy gradient, which is consistent with the commonly accepted inertia effect reported in literature.

## 2. Energy gradient method

In [12-13], it is shown by rigorous derivation that the relative magnitude of the total mechanical energy of fluid particles gained and the energy loss due to viscous friction in a disturbance cycle determines the disturbance amplification or decay. For a given base flow, a stability criterion is written as [12-13],

$$K \frac{v'_m}{u} < Const , \quad K = \frac{(\partial E / \partial n)}{(\partial H / \partial s)} . \tag{1}$$

Here, $K$ is a dimensionless field variable (function) and expresses the ratio of the gradient of the total mechanical energy in the transverse direction and the loss of the total mechanical energy in the streamwise direction for unit volumetric fluid. $E = p + 0.5\rho V^2$ is the total mechanical energy per *unit volumetric fluid*, $s$ is along the streamwise direction and $n$ is along the transverse direction. $H$ is the energy loss per *unit volumetric fluid* along the streamline for finite length. Further, $\rho$ is the fluid density, u is the streamwise velocity of main flow, and $v'_m$ is the amplitude of disturbance velocity.

The so-called energy gradient method showed very good agreement with experiments in three aspects [12-14].

(1) Scaling of the disturbance amplitude required for transition to turbulence with Re. For any type of flows, it has been shown that the magnitude of $K$ is proportional to the global Reynolds number ( $Re = \rho U L / \mu$ ) for a given geometry [13]. Thus, the criterion of Eq.(1) can be rewritten as,

$$Re\frac{v'_m}{U} < Const \quad \text{or} \quad (\frac{v'_m}{U})_c \sim (Re)^{-1} \ . \tag{2}$$

This scaling agrees with the experiments in literature for the pipe flow with transversal injection as the disturbance source, see [13].

(2) The location where the flow instability is first initiated. The value of $K\frac{v'_m}{u}$ in Eq.(1) will reach its critical value in the field with the increase of Re. As is suggested, flow instability is initiated at the place where the criterion is first violated. After the onset of instability at this location, the flow oscillation will become larger and subsequent turbulent transition may be triggered. For a given disturbance level, the distribution of K in base flow determines the critical condition of the flow. From this method, the position of $K_{max}$ is therefore the most dangerous location for initiation of flow instability where the disturbance is greatly amplified. This suggestion concurs with the experiment for plane Poiseuille flow and pipe flow. Nishioka et al's [15] experiments for plane Poiseuille flow showed that the first oscillation of the velocity occurs at y/h=0.50~0.62. The theoretical value of energy gradient method is at y/h=0.58. For the pipe flow, the recent experiments of Nishi et al. [16] showed that the oscillation first started at r/R=0.53-0.73, which is in agreement with our prediction that r/R=0.58 is the most unstable position.

(3) A consistent critical value of Kmax at the minimum Re of turbulence transition.

Experiments showed that when Re is less than a critical value, the flow keeps to its laminar state fairly independent of the magnitude of the disturbance imposed. Dou [12-13] demonstrated that Kmax takes on a consistent value at the condition of transition determined by the experimental data for plane Poiseuille flow, pipe Poiseuille flow as well as the plane Couette flow, with Kmax=370-380.

## 3. Energy gradient function for pipe flow

The Navier-Stokes equation for the fully-developed flow in a pipe of an incompressible Newtonian fluid (neglecting gravity force) in the cylindrical coordinates (r,z) reduces to

$$0 = -\frac{\partial p}{\partial z} + \mu \frac{1}{r} \frac{\partial}{\partial r} \left( r \frac{\partial u_z}{\partial r} \right) \tag{3}$$

where ρ is the fluid density, $u_z$ is the axial velocity, p is the hydrodynamic pressure, and μ is the dynamic viscosity of the fluid. Integrating Eq.(3) using the boundary condition at the solid wall of the pipe, the axial velocity distribution along the radius can be obtained as [12]

$$u_z = u_0 \left[ 1 - \left( \frac{r}{R} \right)^2 \right], \tag{4}$$

where $u_0 = -\frac{1}{4\mu} \frac{\partial p}{\partial z} R^2$. Here, $u_0$ is the maximum velocity at the axis of the pipe, and R is the radius of the pipe.

The ratio of the transversal energy gradient to the streamwise energy loss of unit volume of fluid, K, can then be expressed as,

$$K = \frac{\partial E / \partial n}{\partial E / \partial s} = \frac{\frac{\partial}{\partial r} \left( \frac{1}{2} \rho u_z^2 \right)}{\mu \left( \frac{\partial^2 u_z}{\partial r^2} + \frac{1}{r} \frac{\partial u_z}{\partial r} \right)} = \frac{1}{2} \frac{\rho u_0 R}{\mu} \frac{r}{R} \left[ 1 - \left( \frac{r}{R} \right)^2 \right] = \frac{1}{2} \boldsymbol{Re} \frac{r}{R} \left( 1 - \frac{r^2}{R^2} \right). \tag{5}$$

Here, Re=2ρUR/μ, and U=(1/2)u₀ is the averaged velocity. It can be seen that K is proportional to Re for a fixed point in the flow field.

By letting $\frac{dK}{d(r/R)}=0$, we obtain r/R=0.58. Thus, K has a maximum value at r/R=0.58. The distribution of u, E, and K along the radius for pipe Poiseuille flow is shown in Fig.2. From this result, it is suggested that the flow oscillation at low Re attains its maximum at r/R=0.58 for a given disturbance level.

## 4. Migration of particles in radial direction

When sphere particles are suspended in the flow, they generate drag force by friction and induce pressure drop. Simultaneously, they induce disturbance to the flow due to their motion of presence.  This disturbance can be amplified to a certain extent by the base flow if the Re number is sufficiently large. This kind of disturbance may not be uniform along the radius. The location where the disturbance is large may induce large drag force, and it in turn alters the characteristic of the flow field.

As is well known, for Newtonian flow, one cause of induced secondary flow (transversal flow) can be attributed to the non-uniform distribution of pressure. For dilute suspension flow, to form the Segre-Silberberg annulus, there should be a pressure gradient to drive the particle to reach the annulus in the absence of any other extraneous force. The position of particle concentration should be the place where the pressure assumes its minimum value. What causes this kind of non-uniform pressure distribution? We will discuss below.

In the creeping flow of Newtonian fluid, the variation of pressure comes from the shear stress since the inertia force is zero. In creeping flow of dilute suspension, the additional variation of pressure comes from the shear stress due to the particle motion. However, fluid inertia also plays a role in the said phenomenon of Segre-Silberberg annulus, since in experiment the annulus is not found for Re<<1. What exact role does the inertia play is of the interest here since most authors agreed that it is the fluid inertia which causes the particles to migrate to this location [1-11].

Assuming that (1) The Re of base flow is sufficiently small and the flow is laminar (Re<<2000); (2) The particle suspension is very dilute so that there is no interaction among the particles; (3) The concentration of particle is so small that there is no influence on the velocity distribution, and the velocity distribution keeps its parabolic profile. These assumptions accord with the observed experimental conditions.

Under the above assumptions, the stresses in the flow can be composed of the following three stresses,

$$\tau = \tau_l + \tau_p + \tau_d .\qquad\qquad(6)$$

Here, $\tau_l$ is the shear stress of the Newtonian laminar flow. When the Re is small, $\tau_l$ is not affected by the presence of particles since the particles have no effect on the velocity profile of laminar flow. $\tau_p$ in Eq.(6) is the shear stress induced due to the drag generated by particles which depends on the velocity difference between the particle and the fluid. It is very small in neutrally buoyant spheres suspensions under the above assumptions. Finally, $\tau_d$ is the shear stress induced due to disturbance or oscillation of the velocity resulting from the motion of particles. It is related to the flow Re and concentration. It is just this part of shear stresses that makes the flow behaviour to be distinguished from the flow at other flow conditions.

Following the principle of energy gradient method, the amplification of disturbance depends on the distribution of the energy gradient function. At r/R=0.6, K gets its maximum, and the disturbance is greatly amplified at this position. As such, at r/R=0.6, $\tau_d$ attains its maximum since the particle is also subjected to maximum disturbance/ oscillation. Based on the assumptions, the distributions of $\tau_l$ and $\tau_p$ along the radius are almost the same as that with no disturbance. Therefore, the fluid flow is subjected to maximum drag at the vicinity of r/R=0.6, and the pressure drop dp/dz is correspondingly maximum at r/R=0.6 along the width (Eq.(3)). This will then result in a deficit/kink of pressure at r/R=0.6 in its distribution along the pipe radius (Fig.3). Under this non-uniform distribution of pressure, the particles at the vicinity of r/R=0.6 will tend to migrate to this location, and to be filled in by other particles from elsewhere. Finally, an annulus of concentration of particles is formed at r/R=0.6.

Chao et al [17] carried out simulation on the particle suspension in channel flow. They showed that the concentration annulus of particles will be formed only at low concentration and this annulus occurs at about r/R=0.6, which is in agreement with the experiment in literature. The simulation also showed that spheroid particle concentration induces low pressure region. This result accords with the analysis in this study.

**5. Conclusions**

In this study, it is shown by the energy gradient method that there is a maximum disturbance at r/R=0.58 in dilute suspension flow of sphere particles in a circular pipe. This maximum disturbance allows the particles at this location to assume large amplitude of disturbance and thus leads to large induced shear stress (drag force) at this radius. The large drag force then induces a relatively larger pressure deficit and therefore a kink in the pressure distribution across the width of the pipe. This characteristic feature of pressure distribution is the reason why the particles migrate to the radius of r/R=0.6.

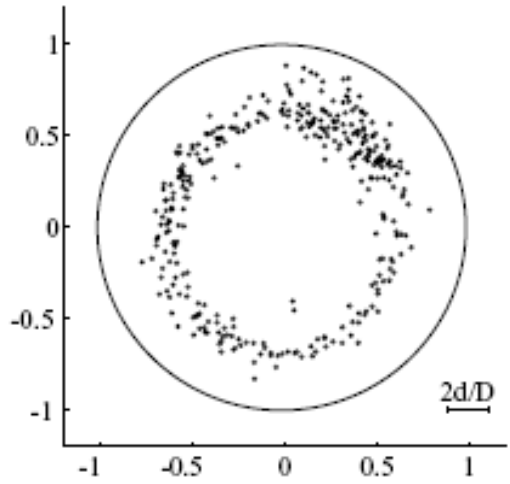

Fig.1 The tubular pinch effect of sphere particles suspended in pipe Poiseuille flow, Re=67 [5].

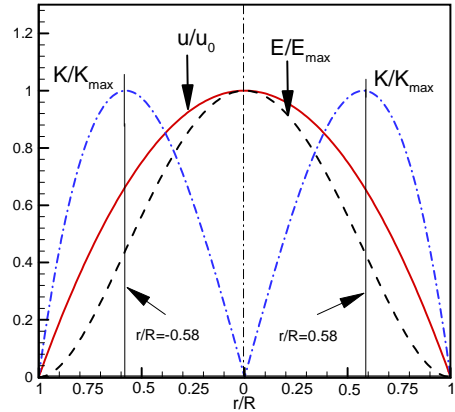

Fig.2 Velocity, energy, and K along versus the radius r/R for pipe Poiseuille flow, which are normalized by their respective maximum.

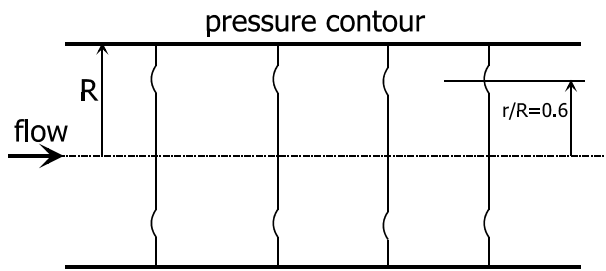

Fig.3 Schematic of contour of pressure in the pipe with sphere particles suspended in the flow.